# Inverse design of glass structure with deep graph neural networks

Qi Wang[1] [*], Longfei Zhang[2]

[1] Science and Technology on Surface Physics and Chemistry Laboratory, P. O. Box 9-35, Jiangyou, Sichuan 621908, China

[2] School of Software, Beihang University, Beijing 100191, China

*Corresponding author. Email: qiwang.mse@gmail.com

## ABSTRACT

Directly manipulating the atomic structure to achieve a specific property is a long pursuit in the field of materials. However, hindered by the disordered, non-prototypical glass structure and the complex interplay between structure and property, such inverse design is dauntingly hard for glasses. Here, combining two cutting-edge techniques, graph neural networks and swap Monte Carlo, we develop a data-driven, property-oriented inverse design route that managed to improve the plastic resistance of Cu-Zr metallic glasses in a controllable way. Swap Monte Carlo, as "sampler", effectively explores the glass landscape, and graph neural networks, with high regression accuracy in predicting the plastic resistance, serves as "decider" to guide the search in configuration space. Via an unconventional strengthening mechanism, a geometrically ultra-stable yet energetically meta-stable state is unraveled, contrary to the common belief that the higher the energy, the lower the plastic resistance. This demonstrates a vast configuration space that can be easily overlooked by conventional atomistic simulations. The data-driven techniques, structural search methods and optimization algorithms consolidate to form a toolbox, paving a new way to the design of glassy materials.

# 1. INTRODUCTION

Tailoring structure for a targeted property is a long-term pursuit of materials[1–8]. This can be formatted as an "inverse design" problem, that is, "given a target property, design the material"[1]. However, such goal is extremely hard for glasses. Without well-defined crystal prototypes to guide the structural search, the disordered and heterogeneous structure of glasses is difficult to manipulate and design. Processing protocols such as quenching rate are usually employed to modify the glass structures, in both simulations and experiments, resulting in some control of glass properties[9]. For example, it is believed that slow quenching reduces the frozen-in excess volume and soft spots in glasses, elevating strength[10]; rejuvenation to a higher-energy state increases deformability, yet in some sacrifice of strength[11]. However, the structural regime visited by such techniques is only a small subset of the astronomic number of glass configurations that can in principle be realized. Recently, swap Monte Carlo (MC) offers an elegant way to search the configuration space, by using energy as the stability measure, and has been successfully used to simulate ultra-slow-quenched polydisperse glasses, even close to the experimental conditions[12]. However, taking energy as the decision metric naturally rule out the vast space of metastable configurations, which, indeed, are very likely to be obtained in experiments[9]. In addition, energy is still an indirect metric to measure the glass properties such as deformation resistance, while these properties are crucial for the practical application of glasses. At present, a property-oriented inverse design protocol, namely directly designing the glass structure to optimize a specific property, is still absent.

To realize such property-oriented inverse design, typically two parts are required: a sampler, which effectively search the configuration space of glasses and proposes new trial configuration, and a decider, or say forward-solver[5], which evaluates whether a trial is acceptable and ideally guides the configuration search. Swap MC and its modified versions can fulfill the role of sampler. However, designing an appropriate decider is dauntingly hard for glasses. The basic premises of the decider are that i) it should be accurate enough in mapping the glass structure to the property of interest, ii) it ideally should be time-efficient, as numerous structural candidates will be evaluated during the design procedure. The advent of machine learning sheds light on this avenue, with successes in correlating the static glass structure with the propensity of plastic deformation[13,14], atomic hopping[15] or thermal activation[16], with computational cost orders of magnitude lower than atomistic simulations. Particularly, the renaissance of deep learning has pushed the limits of prediction accuracy in many domains[17–25] and can serve as a better choice of decider. The powerful parallel computation capability of GPU, in terms of hardware, can also greatly boost the evaluation efficiency, to an unprecedented level.

In this work, combining two state-of-the-art techniques, namely deep graph neural networks (GNNs) as decider and swap MC as sampler, we have realized the inverse design of Cu-Zr metallic glasses (MGs), using the resistance to plastic deformation as an example target property. Graphs are universal models of objects and their pairwise relationships, and numerous structures are be viewed as graphs, including social networks[21], protein interactions[22], organic molecules[23] and crystals[24,25]. GNN is a deep learning-based scheme that operate on graphs[26]. At each layer, information from neighbourhood is aggregated for each node (i.e., atom), and after stacking multiple layers, GNN benefits from the recursive neighborhood expansion, gradually compressing structural information into the low-dimensional atomic embeddings. We note that a recent work from Bapst et al.[20] has achieved very impressive results using GNN in predicting the long-time evolution and deformation susceptibility of Lennard-Jones liquids and glasses. As the application of GNN in the glass research is still in its infancy, there is huge exploration space in the architecture design and application scenarios. Here, in this work, we design a different GNN framework that incorporates multi-head attention, allows for hierarchical structural information extraction in various graph layers, and is perfectly rotationally equivariant by encoding the interatomic distance instead of relative positions. Incorporating GNN and swap MC, we develop an inverse design route that has managed to improve the deformation resistance of typical Cu-Zr glasses with no human intervention. An unconventional strengthening route is uncovered, with a small degree of energy sacrifice, yet achieving a remarkable gain in strengthening. We also design several strategies to accelerate the optimization. To our best knowledge, such property-oriented inverse design of glass structure has not been reported before. This work could open new avenues for the controllable manipulation of glass structure to reach a target property. Besides, the inverse design framework is general and can be conceivably applied to any property of glasses.

## 2. RESULTS

### GNN architecture

Unlike images (i.e., Euclidean graphs), non-Euclidean graphs, like the glass graphs here, are difficult to express and represent. Typically, key processes of GNNs are aggregation, pooling and readout[26]. In each layer, a node aggregates information of itself and its neighbors by graph convolution and pooling, and update its node and bond features for the next layer of message passing. After $N$ layers, each node will aggregate information up to its hop-$N$ neighborhoods. Finally, the node feature vectors can be fed to a readout layer for an upstream task (e.g., node classification or regression).

The schematic picture of our GNN architecture is presented in Fig. 1. Specifically, the graph input is composed of a set of nodes (or vertices) and edges (or bonds), **V** = {$\mathbf{v}_i$}, **E** = {$\mathbf{e}_{ij}$}, where **V** is the set of nodes, $\mathbf{v}_i$ is the initial feature vector of node $i$; **E** is the set of edges, $\mathbf{e}_{ij}$ is the initial edge feature vectors between node $i$ and $j$. Here, the initial $\mathbf{v}_i$ is just a one-hot vector of element types, e.g., [1, 0] for Cu and [0, 1] for Zr, without any hand-tuned features included. The initial $\mathbf{e}_{ij}$ is a vector of Gaussian expanded distance between nodes.

As illustrated in the lower panel of Fig. 1, in each layer, we first update edge feature vectors,

$$\mathbf{e}_{ij}^{(l+1)} = \sigma\,(\mathbf{e}_{ij}^{(l)} + \mathbf{W}_e^{(l)}\,(\,\mathbf{h}_i^{(l)} \parallel \mathbf{h}_j^{(l)} \parallel \mathbf{e}_{ij}^{(l)}\,)) \qquad (1)$$

where $\mathbf{e}_{ij}^{(l)}$ is the edge feature at the $l$-th layer. $\mathbf{W}_e^{(l)}$ is the edge linear transformation's weight matrix. $\mathbf{h}_i^{(l)}$ is the hidden state of node $i$. $\parallel$ denotes vector concatenation. $\sigma$ is an activation function such as sigmoid, softplus, relu or leaky-relu, and we use softplus as default. The residual link[27] is used to connect the output of a layer with its reference input. It makes the layer only learn the additive residual function, which is easier to learn than the original unreferenced function.

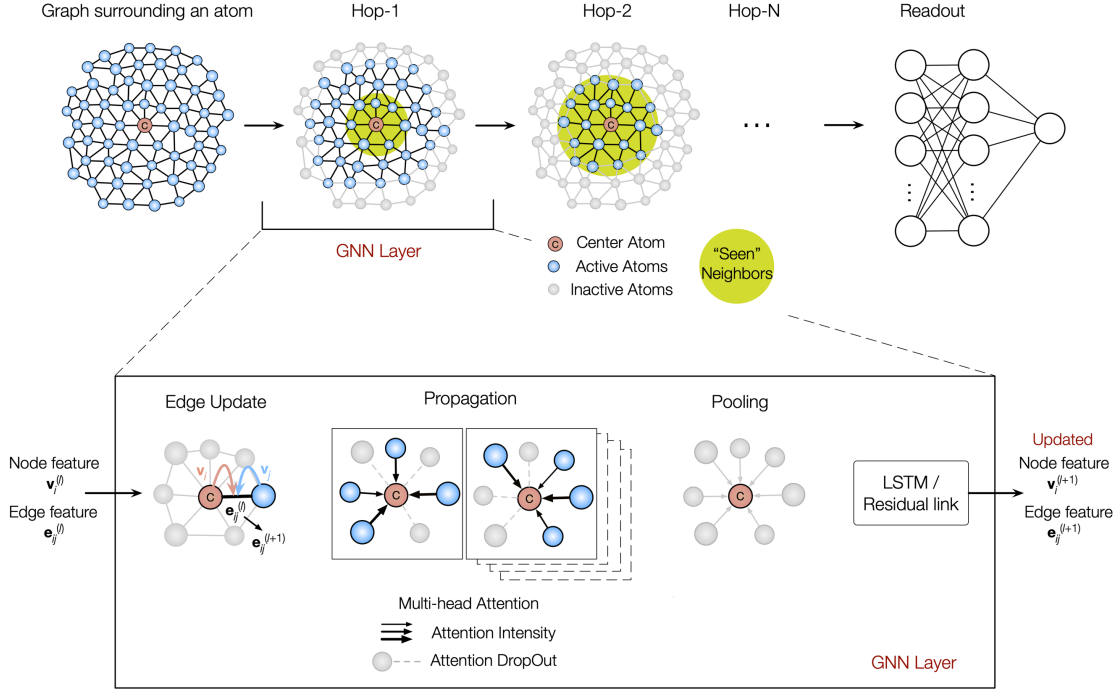

**Figure 1 | Architecture of graph neural networks**. Starting from the glass graph with nodes as atoms and edges as interatomic interactions, the graph networks is stacked with *N* layers, allowing message passing through hop-*N* neighboring layers around each atom. Finally, a readout layer is set up to get the property (regression) or possibility of belong to a class (classification). The processing procedure inside a GNN layer is highlighted in the lower panel.

Next, we update the node hidden state in two steps. First, we calculate each node's local hidden state using a multi-head attention scheme[28,29]. For a *k*-th attention head, we calculate the interactive attention between node *i* and its neighbor *j* and then derive the hidden state between *i* and *j*:

$$a_{ij}^{(l,k)} = \text{Softmax}\,(\mathbf{W}_a^{(l,k)}\,(\,\mathbf{h}_i^{(l)} \,\|\, \mathbf{h}_j^{(l)} \,\|\, \mathbf{e}_{ij}^{(l+1)}\,)) \qquad (2)$$

$$\mathbf{h}_{ij}^{(l+1,k)} = \sigma\,(a_{ij}^{(l,k)} \odot (\mathbf{W}_h^{(l,k)}\,\text{Dropout}(\,\mathbf{h}_i^{(l)} \,\|\, \mathbf{h}_j^{(l)} \,\|\, \mathbf{e}_{ij}^{(l+1)})) \qquad (3)$$

$\mathbf{W}_a^{(l,k)}$ is the weight matrix of the *k*-th attention head. Softmax function normalizes an input vector to a probability distribution, i.e., the attention weights. $\mathbf{W}_h^{(l,k)}$ is the hidden state weight matrix. $\odot$ denotes the element-wise product. We also use the dropout method[30] to let each head focus on a different representation subspace. For each node *i*, we represent all the hidden states, $\mathbf{h}_{ij}^{(l+1,k)}$, between *i* and all its neighbors *j*, from each attention mechanism *k*.

In the second step, we concatenate all heads of hidden states, pool it, and use a "remember" function (abbreviated as RF in Eq. 4) to include the hidden state of the previous layer:

$$h_i^{(l+1)} = \text{RF}\left(\underset{j \in N_i}{\text{Pool}}\left(\underset{k=1}{\overset{K_l}{\|}} h_{ij}^{(l+1,k)}\right)\right) \qquad (4)$$

where $K_l$ is the head number at the $l$-th layer. $\overset{K_l}{\underset{k=1}{\|}}$ means concatenating all heads of $\mathbf{h}_{ij}^{(l+1)-k}$. $N_i$ is the neighborhood of node $i$. Pool indicates the pooling method to get the final hidden state of node. We have tried the set2set[31], mean-pooling, max-pooling and sum-pooling, and sum-pooling is found to be both robust and efficient. We also try three remember functions, namely LSTM[32,33], dense-net[34] and residual link[27], and they are found to perform similarly. In accordance with the edge update function (Eq. 1), we use the residual link at default.

After $N$ layers, the node features will be updated by selectively aggregating structural input from $N$-hop neighborhoods. Finally, we use multi-layer perceptron with dropout as readout layer to do the node regression. These GNN operations are able to operate on graphs of arbitrary shape and graph size. The number of tunable parameters in our GNN model is 131,274. Besides, we also implement the minibatch forward and backward propagation method on graphs[21] to allow for learning large graphs (e.g., containing >$10^5$ atoms) in the future, which would otherwise be easily out of GPU memory. For small-size graphs here (each of 16,000 atoms, to be detailed later), using the entire graph as a minibatch is better. The Pytorch[35] implementation of our GNN model will be made public in GitHub (https://github.com/Qi-max/gnn-for-glass) shortly.

In liquids and glasses, the nearest-neighbors (or say hop-1 neighbors) are frequently studied, in the context of short-range order (SRO)[36]; however, the effects of second-nearest neighbors (hop-2), or say medium-range order (MRO), and beyond are also important but are much harder to encode. GNN offers a natural way to aggregate information from the expanded neighborhood. Meanwhile, our implementation is different from that of Bapst et al.[20] in several aspects: i) we implement multi-head attention to distinguish and quantify neighbor contribution, and the interpretability of GNN is also augmented; ii) we introduce edge dropout to perturb the graph connection in each training epoch, which can be regarded as graph data augmentation and help to improve the robustness of GNN training; iii) instead of the recurrent-style implementation, we let the parameters in the multiple graph layers to vary, allowing for hierarchical structural information extraction in different stages; iv) our GNN framework directly encodes the interatomic distances, which makes it perfectly rotationally equivariant, eliminating the need of data augmentation by adding configurations with varying orientation. In the following, we will demonstrate that our GNN framework achieves remarkable regression accuracy in predicting the plastic susceptibility in Cu-Zr MGs of different compositions and processing histories, and well fulfills the role of a decider in the inverse design experiment.

**GNN predicting atomic-scale plastic susceptibility**

In this work, we train GNN models to predict the plastic propensity of atoms, from the static, unstrained structure. The models obtained will then serve as decider for the later design of glasses that are ultra-resistant to plastic rearrangement. As we'd like to the glass to be strengthened in a global manner, rather than merely be strengthened under a specific loading mode or direction, we lead our GNN to learn to predict a more comprehensive deformation propensity. We stimulate twelve athermal quasi-static (AQS)[37] loading conditions and average the $\ln(D_{\min}^2)$ at each strain as the deformation indicator, i.e., the regression target of GNN (simulation details in Methods). We quench 120 configurations, each of 16,000 atoms, of $Cu_{50}Zr_{50}$ and $Cu_{64}Zr_{36}$ under a rate of $10^{10}$ Ks$^{-1}$. The configurations of each glass are divided into three parts: 100 (i.e., a total of 1.6 million atomic environments) for training, 10 configurations for validation and 10 configurations for test. GNN models are trained on the training configurations, and models with the lowest validation error is selected. Finally, the test score is derived (note that the test set is totally unseen and does not participate in model selection). We also simulate 10 $Cu_{50}Zr_{50}$ and $Cu_{64}Zr_{36}$ configurations under a slower quenching rate of $10^9$ Ks$^{-1}$, to test the generalizability of our GNN models between quenching rates or/and compositions. Pearson correlation coefficient is used as the major scoring metric (also for ease of comparison with previous works). We allow the GNN model to train for ~5,000 epochs, and in practice, the model usually reaches the best validation score at ~1,000 epochs.

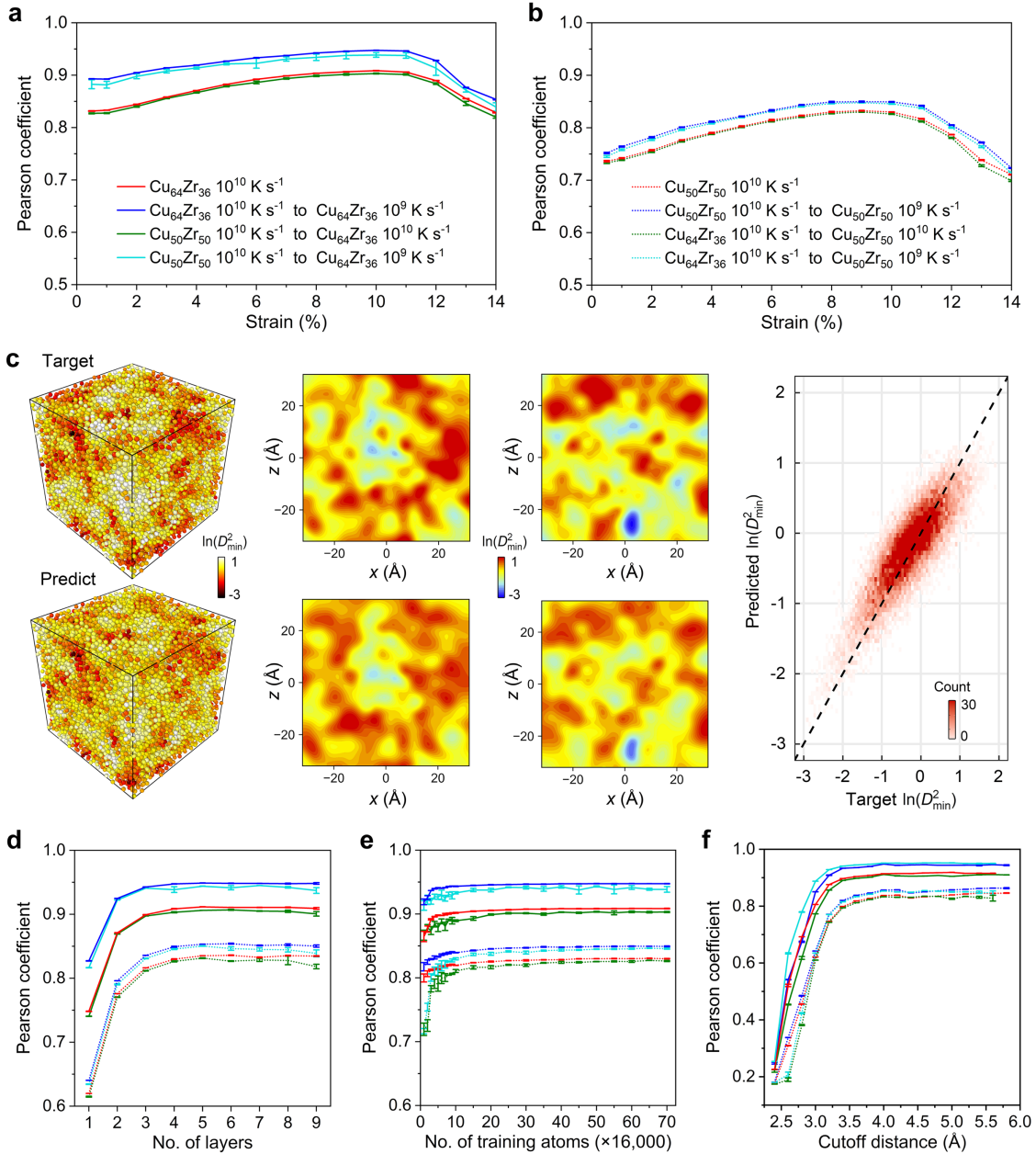

**Figure 2 | Predicting the plastic propensity using GNN.** (**a-b**) Pearson coefficient of $\ln(D^2_{min})$ prediction for each strain of (**a**) $Cu_{64}Zr_{36}$ and (**b**) $Cu_{50}Zr_{50}$, respectively. GNN generalization results between quenching rates or/and compositions are also presented. Each point and the error bar show the average and standard deviation of an ensemble of ten independently trained GNN models. (**c**) Target (simulated, upper row) versus GNN predicted (lower row) $\ln(D^2_{min})$ of a $Cu_{50}Zr_{50} - 10^9$ K s$^{-1}$ configuration at strain 10%. Linear interpolation of $\ln(D^2_{min})$ in typical slices of 3 Å thickness are compared in the middle panel (more slices of different strains, compositions and quenching rates in Supplementary Figs. 1-4). Parity plot of the target and predicted $\ln(D^2_{min})$ is in the right panel. (**d-f**) Pearson coefficient with varying (**d**) number of GNN layers, (**e**) number of atoms (configurations, each containing 16,000 atoms) for training, and (**f**) cutoff distance in designating nearest (hop-1) neighbors when building the graphs around atoms.

Figures 2a and 2b show the Pearson coefficients of $\ln(D^2_{min})$ prediction for strain 0.5% to 14% of $Cu_{64}Zr_{36}$ and $Cu_{50}Zr_{50}$. We find that the quality of GNN prediction keeps increasing until the strain of 10%, in vicinity to the yielding point of glass. This trend is similar to that of Bapst et al.[20], which employs an "incremental" prediction scheme, i.e., using the configuration at a given tilt to predict the displacement as the tilt increases by 4%[20]. Here, we managed to use the undeformed structure to predict the deformation heterogeneity at large strains. For closer scrutinization, we show the target versus GNN predicted $\ln(D^2_{min})$ of a $Cu_{50}Zr_{50} - 10^9$ K s$^{-1}$ configuration at strain 10% (Fig. 2c). The prediction is derived by GNN model trained and verified in $Cu_{50}Zr_{50} - 10^{10}$ K s$^{-1}$ configurations (i.e., generalization between quenching rates). Satisfactory correspondence can be seen from the three-dimensional $\ln(D^2_{min})$ distribution (left), the linear interpolation of $\ln(D^2_{min})$ in typical slices of 3 Å thickness (middle, please find more slices of different strains, compositions and quenching rates in Supplementary Figs. 1-4), as well as from the parity plot (right). To further estimate the reliability of prediction, we independently train an ensemble of ten GNN models (bagging), and find that their test scores are quite close, suggesting that our training is very stable (Figs. 2a and 2b). We also test using the predicted $\ln(D^2_{min})$ to classify the soft or hard atoms at various strains and achieve remarkable accuracy (Supplementary Figs. 5-8). As an example, in the soft end, the classification metric, area under receiver operation characteristic curve (AUC-ROC), of $Cu_{50}Zr_{50} - 10^9$ K s$^{-1}$ reaches ~0.968 under a fraction threshold $f_{thres}$ of 0.5% and ~0.924 upon larger $f_{thres}$ of 10% before yielding (Supplementary Fig. 7a). The classification performance in the hard end is even better, with AUC-ROC > 0.94 for all strains before yielding (Supplementary Fig. 7b). For $Cu_{64}Zr_{36} - 10^9$ K s$^{-1}$, as another example, the AUC-ROC for soft atoms reaches ~0.983 ($f_{thres}$ = 0.5%) and ~0.955 ($f_{thres}$ = 10%), and the AUC-ROC for hard atoms steadily > 0.98 before yielding (Supplementary Fig. 5). Besides, it is found that GNN models fitted in different compositions also generalize well (Figs. 2a and 2b), manifesting the power of GNN in representing the heterogenous atomic environments present in different glasses.

We also investigate the performance of GNN as a function of the number of graph layers ($N_{layer}$), training atoms ($N_{atom}$), and cutoff in defining nearest-neighbors ($R_c$) (Figs. 2d-2f). The GNN performance gradually stabilizes at $N_{layer}$ of 4 (Fig. 2d), suggesting that information up to 4$^{th}$ neighbor shell is of critical importance to the prediction. The score experiences a rapid elevation with the increase of $N_{atom}$, and even ~80,000 atomic environments (equivalently, 5 configurations, each containing 16,000 atoms) are able to yield satisfactory results (Fig. 2e). The low requirement of $N_{atom}$ indicates an impressive learning efficiency of our GNN framework. For $R_c$, ~4.0 Å can yield optimal results (Fig. 2f) and avoid neighborhood explosion with too large cutoff. With the highly accurate GNN models at hand, we will proceed to demonstrate the feasibility of inverse design by integrating GNN and swap MC.

# Inverse design of deformation-resistant glass structure

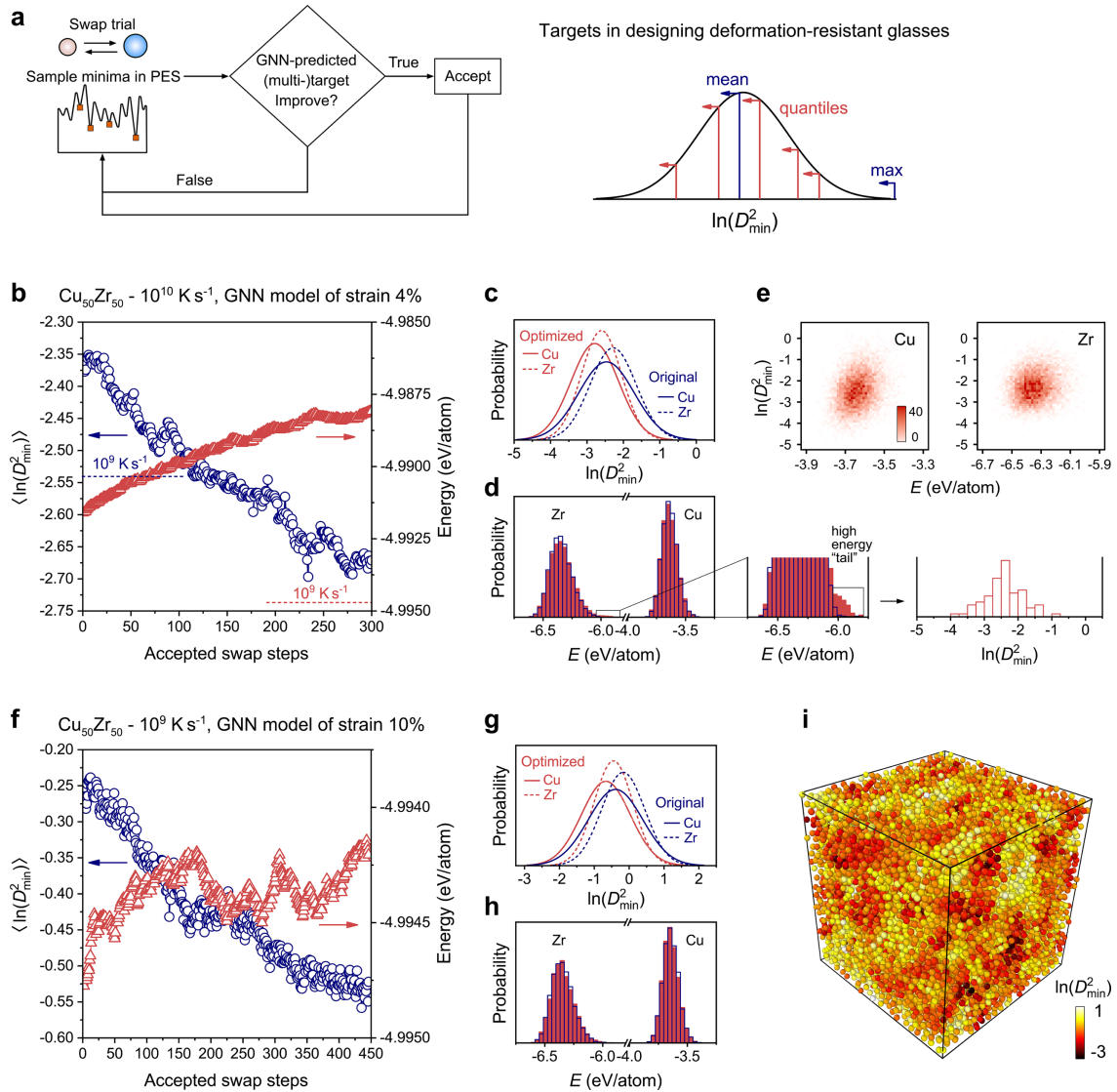

**Figure 3 | Inverse design of deformation-resistant glass structure**. (**a**) Schematic of the iterative design scheme, based on swap MC as sampler and GNN as decider. The right panel shows the targets to be optimized in designing the deformation-resistant glasses. (**b**) Evolution of configuration-average $\langle\ln(D_{min}^2)\rangle$ and energy, when using the GNN model for strain 4% to optimize a $Cu_{50}Zr_{50} - 10^{10}$ K s$^{-1}$ configuration. (**c**) Distribution of $\ln(D_{min}^2)$ of Cu and Zr atoms before and after optimization. (**d**) Distribution of energy before (blue, open histograms) and after (red, filled histograms) optimization. A high energy "tail" in the energy spectra of Zr atoms are highlighted and the $\ln(D_{min}^2)$ distribution of these high-energy atoms are presented in the right panel. (**e**) Weak correlation between local energy and $\ln(D_{min}^2)$. (**f**) Evolution of configuration-average $\langle\ln(D_{min}^2)\rangle$ and energy when using the GNN model for strain 10% to optimize a $Cu_{50}Zr_{50} - 10^9$ K s$^{-1}$ configuration. (**g**) The distribution of $\ln(D_{min}^2)$ before and after optimization. (**h**) The distribution of energy before and after optimization. (**i**) The simulated $\ln(D_{min}^2)$ distribution of the optimized configuration. A systematic decrease of $\ln(D_{min}^2)$ can be observed with comparison to the original configuration ("Target", upper left in Fig. 2c).

We develop a GNN-guided, property-oriented swap MC route to strengthen model $Cu_{50}Zr_{50}$ configurations and improve their resistance to plastic rearrangement. A schematic is presented in Fig. 3a and the detailed procedure is presented in Methods. We first take a $Cu_{50}Zr_{50} - 10^{10}$ K s$^{-1}$ configuration at starting configuration and the GNN model for strain 4% as decider (Fig. 3b). We tried 30000 rounds of trials, and ~1.2% were accepted. To validate the optimization, we exhaustively simulate the AQS deformations for each accepted configuration and calculate the $\ln(D^2_{min})$ at strain 4% (Fig. 3b). After ~120 accepted steps, the $\ln(D^2_{min})$ of the optimized $Cu_{50}Zr_{50} - 10^{10}$ K s$^{-1}$ configuration is already lower than that of the $Cu_{50}Zr_{50} - 10^9$ K s$^{-1}$ glass (as highlighted by the blue dashed line in Fig. 3b), which is the lowest quenching rate used in this work. The detailed $\ln(D^2_{min})$ distributions of Cu and Zr before and after optimization are shown in Fig. 3c. The plastic susceptibility of Cu and Zr both remarkably decreases after optimization. We note that to reduce the computation overhead and accelerate optimization, we select the GNN model with the highest validation score as decider, rather than averaging the predictions from the ensemble of GNN models (whose scores are indeed similar, as observed in Figs. 2a and 2b).

Interestingly, if we look at the energy, the energy indeed increases during optimization (Fig. 3b). This indicates that the optimized state is not an energetic ground state, yet stands out as a "geometrically ultra-stable" state. In experiments, locally (even not globally) stable configurations can still be manufactured even if they do not correspond to the thermodynamic ground state. Furthermore, despite the energy increases, the increase is merely ~3.75 meV/atom, suggesting that this state has a good chance to be sampled in real experiments, even from the energetic perspective (experimental rejuvenation can introduce ~300 meV/atom into the glass[11]). The optimized configuration has a slightly wider energy dispersion (Fig. 3d). A high energy tail appears in the energy spectrum of Zr; however, their plastic inclination (right panel in Fig. 3d) is close to the overall distribution of Zr (Fig. 3c). This is further supported by the weak correlation between local energy and $\ln(D^2_{min})$, with low Pearson correlation coefficients of 0.269 and 0.148 for Cu and Zr, respectively (Fig. 3e).

We further try optimizing the plastic resistance up to a larger strain of 10%, starting with a $Cu_{50}Zr_{50} - 10^9$ K s$^{-1}$ configuration (Fig. 3f). The GNN model for strain 10% is used as decider. We tried 90000 rounds of swap-displacement trials, and ~0.5% were accepted. After optimization, the $\ln(D^2_{min})$ of Cu and Zr also markedly shifts to the lower end (Fig. 3g). In this case, the energy only shows a marginal increase of ~0.55 meV/atom (Figs. 3f and 3h), suggesting that energetic stability is more important for the long-term plastic resistance than short-term resistance. The spatial distribution of $\ln(D^2_{min})$ at strain 10% is derived for the optimized configuration (Fig. 3i), which has been systematically strengthened from the original state (upper left in Fig. 2c).

Given the higher energy (despite not much) of the "geometrically ultrastable" configurations, it is of interest to check their stability under finite temperatures and examine whether the enhanced plastic resistance can be maintained. To this end, we first anneal the two optimized configurations (Fig. 3c) at the room temperature of 300 K (Supplementary Fig. 9). The ensemble is set to be canonical (NVT) to avoid the effect of volume change on plastic resistance. The introduction of temperature leads to a notable energy increase, and the system is allowed to equilibrate at 300 K for 2 ns (Supplementary Figs. 9a and 9b). After that, we reapply the conjugate-gradient method to get the inherent structure after annealing (this step is also essential for the later AQS deformation to examine the plastic resistance, as AQS requires the inherent structure as starting structure). Afterwards, we perform AQS deformation simulations under the same conditions, and the distribution of $\ln(D_{\min}^2)$ for the GNN-optimized configuration, the GNN-optimized configuration after annealing at 300 K, and the original unoptimized configuration are organized in the Supplementary Figs. 9a and 9b. It is found that the after annealing at 300 K, the optimized plastic resistance can still be maintained, with a systematic strengthening from the original state. The stability of the GNN-optimized $10^9$ K/s configuration is relatively stronger than the $10^{10}$ K/s one. Thus, we further try annealing the $10^9$ K/s configuration at a much higher annealing temperature of 600 K, which is only ~150 K below the glass transition temperature ($T_g$). It turns out the plastic resistance can still be maintained well (lower panel in Supplementary Fig. 9b). This can be understood in terms of the large energy barrier between the local minima of PES. Despite the optimized state has a minor energy increase from the original state, the energy barriers between these structures are indeed much larger. This enables the optimized state to well resist the imposed thermodynamic stimulus without pronounced structural change.

When the melt is quenched from high temperature, due to the drastic dynamical slowdown, it is very likely to be confined in one of the local minima in PES, rather than the global minimum. In this work, we present an interesting data-centric, property-oriented protocol to explore the glass PES. Results indicate that even within a very small energy window and at a fixed composition, the property of glasses has a large room for adjustment. This can provide some theoretical basis for explaining the varying glass property in experiments, even under similar processing routes, and encourages further exploration of the vast and intriguing structural space of glass materials.

We believe that the optimization can go further, if we continue this optimization procedure. But it will take extra time. Our main purpose here is to demonstrate that the collaboration of GNN and swap MC is a powerful way to meet the inverse design challenge we are dealing with. Here we reveal that GNN well serves as surrogate model to perform tests on "virtual geometries" and controllably explore the configuration space in search of structures with desired properties. The high accuracy of GNN makes it possible to replace the expensive simulations for preselecting and optimizing the glass structures. As our GNN predicts the plastic resistance at the atomic-scale, we can even directionally strengthen the "weak spots" of glass structure instead of conducting completely stochastic trials over all atoms (to be elaborated in the Discussion section). Besides, the optimization task here is to improve the plastic resistance, and one can devise different optimization tasks by developing customized GNN models and designing customized optimization metrics. One can even use this strategy to design complex glass structures with specific pattern of hard and soft regions for more sophisticated use-cases.

## Interpreting the unconventional strengthening

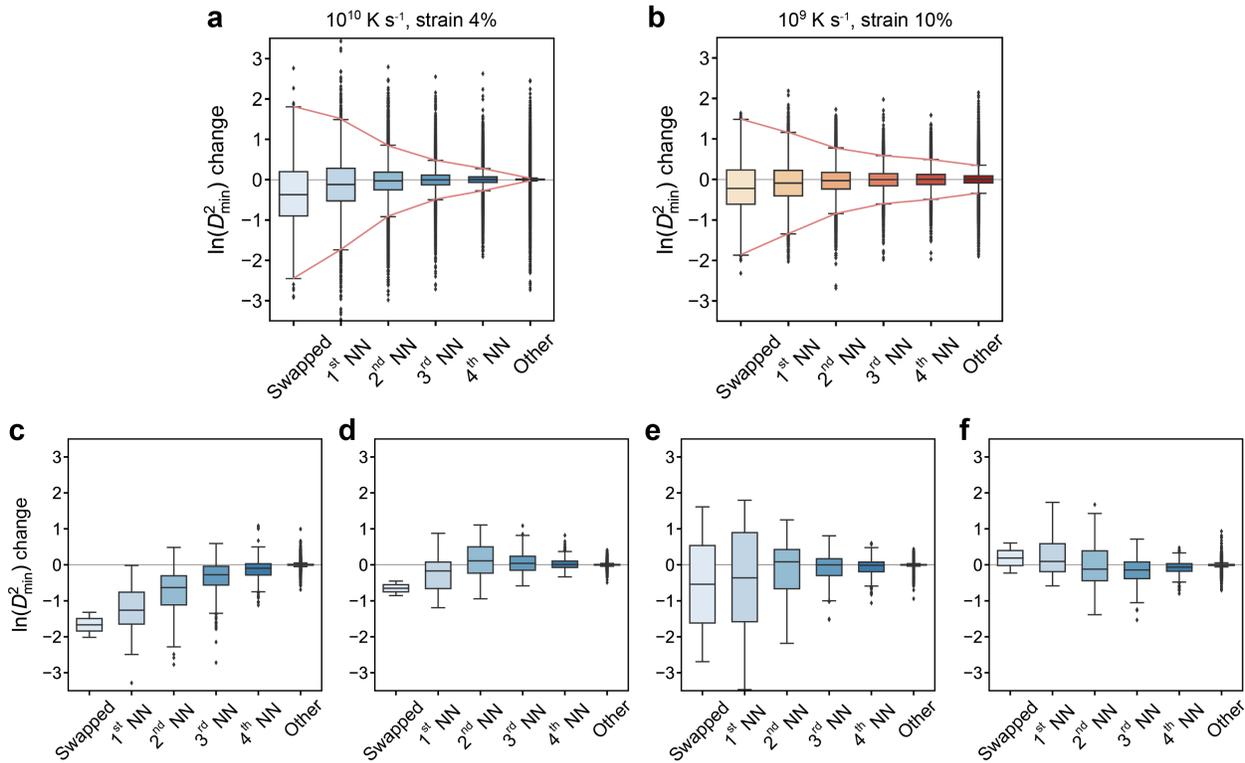

**Figure 4 | The strengthening mechanism**. (**a-b**) Statistics of $\ln(D^2_{min})$ change after each optimization step when using (**a**) GNN model for strain 4% to strengthen $Cu_{50}Zr_{50}$ – $10^{10}$ K s$^{-1}$ and (**b**) GNN for strain 10% to strengthen $Cu_{50}Zr_{50}$ – $10^9$ K s$^{-1}$. Atoms are grouped into six categories: the swapped atoms, 1$^{st}$, 2$^{nd}$, 3$^{rd}$ and 4$^{th}$ neighbors of the swapped atoms, and all other atoms. All the accepted steps are included in the statistics. In the box plots, ends of box spans from 25 to 75% percentile, black line in box represents median, whiskers show 1.5 times the inter-quartile range, and points outside the whiskers show outliers. (**c-f**) Typical $\ln(D^2_{min})$ change after a single optimization step.

It is interesting to take a closer look at how the strengthening works. We extract the statistics of $\ln(D^2_{min})$ change after each accepted optimization step (Figs. 4a and 4b). Atoms are grouped into six categories: the swapped atoms (i.e., two in each step), 1$^{st}$, 2$^{nd}$, 3$^{rd}$ and 4$^{th}$ neighboring shells of the swapped atoms, and all other atoms farther away. Results show that the optimization mainly focus on the swapped atoms and their 1$^{st}$ neighbors; the farther from the swapped atoms, the weaker the optimization effect. Figures 4c-4f present some representative scenarios in a single optimization step: i) the swapped atoms and their neighbor regions up to the 4$^{th}$ shell are all notably strengthened; ii) the swapped atoms and their 1$^{st}$ neighbors are strengthened, while neighbors beyond 1$^{st}$ shell are softened slightly; iii) a compromise case where one swapped atom and some of the neighbors are stabilized, whereas the others are softened; iv) the swapped atoms and their 1$^{st}$ neighbors are softened yet the 2$^{nd}$, 3$^{rd}$ and 4$^{th}$ neighbors are strengthened. These scenarios have indicated the influence of high-impact atoms (or atom pairs) for strengthening. After accumulating swaps of these high-impact atoms, the plastic resistance can be enhanced in a controllable way.

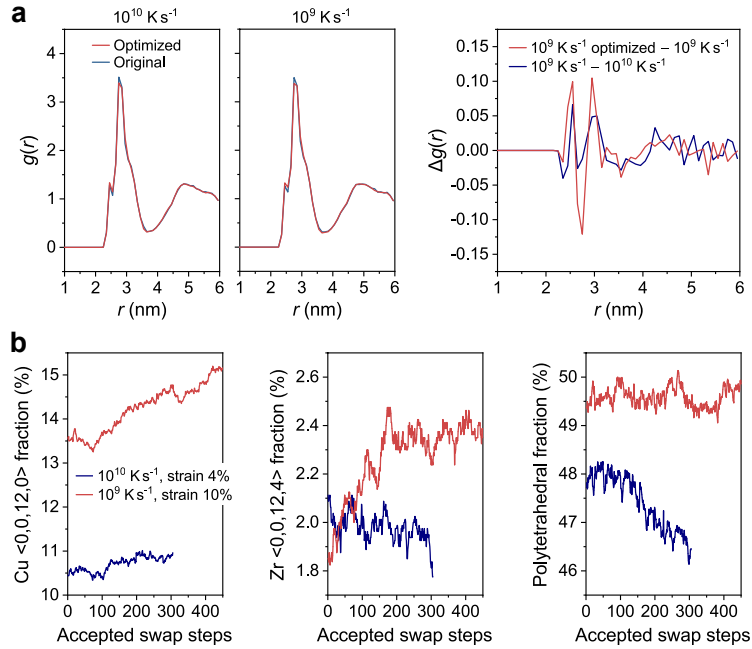

**Figure 5 | Structural evolution with optimization**. (**a**) Pair correlation function, $g(r)$, of the original $Cu_{50}Zr_{50}$ $10^{10}$ K s$^{-1}$ and $10^9$ K s$^{-1}$ configurations and of the GNN-optimized configurations. $\Delta g(r)$, the difference of $g(r)$ between two states, induced by the GNN-guided optimization and the decrease of quenching rate are derived and compared. (**b**) Tracking the fraction of typical Voronoi motifs, i.e., Cu-centered <0, 0, 12, 0> icosahedra, Zr-centered <0, 0, 12, 4> and polytetrahedral clusters, during the optimization process.

Given the success of GNN-guided optimization, it is of interest to take a sip of the structural evolution during optimization. As a baseline, we first derive the pair correlation function, $g(r)$, of the original and optimized states (Fig. 5a). We find the $g(r)$ values before and after the optimization are overall similar. To examine the subtle changes of peak heights, we further extract the $\Delta g(r)$, the difference of $g(r)$ between the two states of interest, $g_1(r) - g_2(r)$. It is known that the GNN-guided optimization as well as the decrease of quenching rate can both enhance the plastic resistance, so we extract the $\Delta g(r)$ caused by the two protocols, as a comparison (right panel of Fig. 5a). Interestingly, for the two cases, the "ups and downs" with $r < 3.2$ Å show some similar traits. This $r$ regime is mainly dominated by Cu-Cu and Cu-Zr pairs, which suggests that the bonding between Cu-Cu and Cu-Zr may show some similar changes in distance. At larger $r$ beyond 4.0 Å, which falls in the 2$^{nd}$ peak envelope of $g(r)$, the change in the two cases is basically opposite (e.g., one is up and another is down). This may indicate a different mechanism in this $r$ regime, dominated by MRO and beyond (more analysis will be conducted in the future work).

As another analysis, we further perform Voronoi tessellation analysis, a common structural analysis framework for glasses, to track the structural changes during optimization. Figure 5b monitors some typical Voronoi motifs that have long been proposed to be critical to glass stability[36]. During optimization, the fraction of Cu atoms centered by <0, 0, 12, 0> icosahedra increases by ~1.5% and ~0.5% for $Cu_{50}Zr_{50}$ – $10^{10}$ K s$^{-1}$ and $10^9$ K s$^{-1}$, respectively (Fig. 5b). This increase, though insignificant, seems to agree with the long-lived understanding of more icosahedra, more stable glass. However, the increase of icosahedra is far less prominent compared to the $\ln(D^2_{min})$ change during optimization. For example, after 120 swaps, the $\ln(D^2_{min})$ of $Cu_{50}Zr_{50}$ – $10^{10}$ K s$^{-1}$ is already lower than that of $Cu_{50}Zr_{50}$ – $10^9$ K s$^{-1}$ (Fig. 3b); however, the corresponding icosahedra fraction at that step is still ~3% lower than that of $10^9$ K s$^{-1}$ (Fig. 5b). For the Zr atoms surrounded by <0, 0, 12, 4>, the fraction experiences a greater oscillation and overall shows a minimal change, i.e., increases by ~0.6% and decreases by ~0.3% for $Cu_{50}Zr_{50}$ – $10^{10}$ K s$^{-1}$ and $10^9$ K s$^{-1}$, respectively (Fig. 5b). Similar oscillation is observed for the polytetrahedral clusters[36] (with $2n_4 + n_5 = 12$, where $n_4$ and $n_5$ are 4-fold and 5-fold Voronoi indices, Fig. 5b). These results indicate a weak correlation between plastic resistance and these representative clusters.

Resorting to the physical-inspired features may help us gain some intuitive understanding of the structural reasons for the improved plastic resistance, however, the connection is still largely empirical. We are working on clarifying the structural mechanism of strengthening by directly interpreting the GNN models, such as quantifying the neighbor contributions based on attention weights and mining the graphic traits of high-impact atoms or substructures for plastic resistance. The aim is to make the GNN models less like a "black box", which can then provide further details in addition to the above structural analysis.

## 3. DISCUSSION

This work demonstrates that the collaboration of GNN as decider and swap MC as sampler open an avenue for the property-oriented inverse design of glass structure. In order to achieve this goal, we also design several strategies to accelerate the optimization:

(i) Instead of randomly selecting Cu and Zr for swap trial, we use the GNN-predicted plastic susceptibility to bias the selection. The probability for an atom $i$ to be selected is,

$$p_i = \frac{\text{softmax}(z_i)^n}{\sum_A \text{softmax}(z_i)^n} \quad (5)$$

where $z_i$ is the GNN-predicted $\ln(D^2_{min})$ of atom $i$, A is the element type of atom $i$, and n is an exponential number controlling the inclination of selecting large $\ln(D^2_{min})$ atoms (n is set as 6 in practice). In each trial, we will calculate $p_i$ for each Cu and Zr atom, and sample one Cu atom and one Zr atom according to the probability distribution of each element type. This allows us to "directionally strengthen" the soft spots without changing too much of the hard backbone.

(ii) We focus on sampling the inherent structures, namely local minima in the PES, to reduce the search space. In practice, after an atom pair swap, we apply the conjugate-gradient algorithm to get the inherent structure and let GNN decide whether to accept or not based on the inherent structure. Thus, the structural search approach is more of a molecular static method instead of a molecular dynamical one. In typical swap MC simulations, whether to accept a swap trial is normally determined by the instant gain after swapping; however, when the size difference of atoms is too large, the acceptance rate of swap trials will quickly diminish to ~0. By allowing the environments to modify to accommodate the swapped atoms, we can also make more feasible decision of whether this swap is acceptable.

(iii) After each trial, we do not have to reconstruct the entire graph. The swap can be realized by exchanging the atom features of the selected atom pair. After obtaining the inherent structure, we only re-determine the neighborhood for the hop-1 and hop-2 neighbors of the swapped atoms, and then reconstruct graphs for those atoms. After that, we will recalculate the properties of the atoms up to hop-$N_{layer}$ around those atoms with graph reconstructed, as the "affected zone" of graph reconstruction extends to hop-$N_{layer}$ neighbors. For small systems, however, such affected zone can easily propagate to the whole system; but this can be useful in future study of large systems, helping to avoid repeated calculations.

(iv) Finally, in our implementation, the entire swap trial and GNN evaluation procedure is executed in GPUs, so the computation speed is fast. The GNN takes significantly longer for training, but is much faster in terms of prediction (even in milliseconds). The unprecedented parallel computation power of GPU greatly accelerates the optimization.

The possible performance degradation of the GNN in the optimization procedure is an issue that needs caution. This is because the distribution of target variable (or in essence, the atomic environments) can gradually get away from the training distribution. In this work, the change of Pearson coefficient on the accepted configurations is monitored in Supplementary Fig. 10. The Pearson coefficient decreases from 0.809 to 0.741 for $10^{10}$ K s$^{-1}$, strain 4% and from 0.851 to 0.821 for $10^9$ K s$^{-1}$, strain 10%. While we conclude that the accuracy is still sufficient to support the optimization, and the optimization continues to make progress (as seen in Figs. 3b and 3f), it would be of interest to incorporate other ideas such as active learning to iteratively improve the GNN model. For example, one can set some checkpoints to validate the optimization performance by conducting AQS simulations for the accepted configurations, and if there is a large performance loss, the configurations can be added to the training set and retrain the GNN model.

It is also of interest to try using the true AQS simulations directly as decider, which could form some upper bound for the optimization performance. Replacing the GNN estimation with a true MD simulation will undoubtedly leads to an improved accuracy in decision making. Despite GNN has achieved high Pearson coefficients (Fig. 2), there are still discrepancies with the true MD simulation. This has caused fluctuations of $\langle \ln(D_{min}^2) \rangle$ for the configurations accepted by GNN as decider (Figs. 3b and 3f). $\langle \ln(D_{min}^2) \rangle$ should monotonically decrease if we elaborately simulate the deformation behaviors of each trial configurations and then decide whether to accept or not. However, this is time-consuming. If we simulate 12 different modes of AQS deformations, the computation time (e.g., using Intel Xeon Gold 6126 CPU @ 2.60GHz) is roughly 100 CPU hours. If we employ GNN to estimate, the prediction time (e.g., using NVDIA Tesla P100) is only 0.17 seconds. The time difference will be quite large if we do ~10000 rounds of trial configurations (as Cu and Zr have a large size difference, the acceptance rate is relatively low). This makes the GNN a reasonable and faster surrogate model for the MD simulations.

Furthermore, despite the prediction time of GNN is sufficiently short for the current glass system, the tradeoff between speed and accuracy is an issue worth noting. In certain cases, it can be possible to build much faster networks with a small decrease in accuracy, and the optimization, as an example, can run more rounds in a fixed computing time. In addition, when designing the model infrastructure, especially when deciding between the "massive" and "lightweight" models, issues such as the training time, the requirement of memory as well as the inclination of overfitting should also be taken into consideration. Currently, it is hard to figure out where the optimal tradeoff lies, which requires further studies towards this direction.

Taken together, this work forms a first attempt to introduce artificial intelligence methods to tackle the inverse-design problem in the domain of glasses. This is achieved by pairing the cutting-edge deep learning framework, GNN, as surrogate model, with tailored structural search methods and optimization criteria. The high-accuracy of our GNN model and massive parallel computation power of GPU solve the most time-consuming part of the structural search, significantly reducing the time for evaluating the numerous glass candidates. The protocol proposed here paves a new way to design tailored glass structures with particular desired functionalities.

We will explore more optimization algorithms, such as evolutionary algorithm, to see if a global optimal structure could be obtained more efficiently. We also plan to try generator networks, based on conditional variational autoencoders or generative adversarial networks, to test whether such networks can reliably propose promising structures in "one-shot" based on latent vector and target property. It is also interesting to combine multiple glass properties into a multi-objective optimization task. In addition, the current case is designing glass configurations at a fixed stoichiometry, we will also explore to allow the stoichiometry to vary, thereby searching over a larger compositional space. With the rigorous development of artificial intelligence in recent years, data-driven methods and optimization algorithms can now be integrated to form a toolbox, revealing the rules behind complex phenomena. Applying this toolbox to glass research, we can deepen the understanding of key scientific issues in the field of glasses, such as structure-property correlation, and provide theoretical basis for the reverse regulation of glass properties.

## Methods

### Glass samples preparation

Molecular dynamics (MD) simulations using LAMMPS[38] are employed to prepare the Cu-Zr glass models, using a set of embedded-atom-method (EAM) potentials[39]. $Cu_{64}Zr_{36}$ and $Cu_{50}Zr_{50}$ samples each containing 16,000 atoms are quenched to 0 K from melts above the melting points. The quenching is performed at a rate of $10^9$ or $10^{10}$ K s$^{-1}$, using a Nose-Hoover thermostat with zero external pressure. Periodic boundary conditions (PBC) are applied in all three directions during MD simulation. The timestep is 1 fs.

### AQS data generation

For each glass configuration, we stimulate twelve AQS[37] loading conditions, i.e., uniaxial tension and compression along x, y and z direction (6 conditions) and simple shear along xy+, xy-, yz+, yz-, xz+ and xz- (6 conditions). On each deformation step, an affine strain of $10^{-4}$ is imposed along the loading direction, followed by energy minimization using conjugate-gradient method. The simulations are conducted using LAMMPS[38] and periodic boundary conditions (PBC) were applied in all three directions.

The plastic indicator, namely the natural logarithm of non-affine displacement ($D^2_{min}$)[39], is calculated for strain 0.5% to 14%, to quantify the deformation propensity of atoms at each strain. The use of $\ln(D^2_{min})$ as the plastic indicator is to convert the "long-tail", lognormal-like distribution of $D^2_{min}$ to a normal-like distribution for ease of training. In addition, more complex deformation modes such as pure shear are not included, as the twelve modes can already reflect a comprehensive plastic resistance of each atom, while more complex modes could be considered as superposition of the elementary modes.

**Swap Monte Carlo**

Swap MC are recently proposed as a powerful method to search the configuration space of the disordered materials[12]. In swap MC simulations, two types of trials are conducted, i) atom pair swap and ii) displacement move. The Metropolis–Hastings criterion[40], based on the energy difference and temperature, is usually used to decide whether a certain trial is acceptable,

$$p = \begin{cases} 1 & \text{if } \Delta E \leq 0 \\ e^{-\frac{\Delta E}{kT}} & \text{if } \Delta E > 0 \end{cases} \quad (6)$$

where $p$ is the acceptance probability, $\Delta E$ is the energy difference between the trial state (with an element pair swapped) and old state, $E^{trial} - E^{old}$; $k$ is the Boltzmann constant and $T$ is the simulation temperature.

**GNN-guided, property-oriented swap Monte Carlo**

In this work, we let the GNN predictions to replace the role of energy ($E$) in Eq. 6 and design a few rules to decide whether to accept a trial (as illustrated in Fig. 3a). The procedure of the GNN-guided, property-oriented swap MC is as follows,

(i) Apply the GNN to predict the softness metric, $\ln(D^2_{min})$, of each atom in the starting configuration, and record the initial values of the designed target functions. As an example, the right panel in Fig. 3a shows the seven target functions (the maximum, mean and quantiles) we have used in this work;

(ii) Select a pair of atoms of different species according to probabilities based on $\ln(D^2_{min})$ (see the point (i) in Discussion and Eq. 5 for more details) and swap their species, and then obtain the inherent structure using conjugate-gradient method. The volume is kept fixed to avoid the volumetric effect on plastic resistance. The acquisition of inherent structure is similar to that applied in AQS deformation (Methods). Re-calculate the $\ln(D^2_{min})$ for those atoms that are affected by the swap trial (see the point (iii) in Discussion). Afterwards, for each target function, calculate its change to determine the associated acceptance probability,

$$p = \begin{cases} 1 & \text{if } \Delta F \leq 0 \\ e^{-\frac{\Delta F}{C}} & \text{if } \Delta F > 0 \end{cases} \quad (7)$$

where $\Delta F$ is the target function difference between the trial state and old state, $F^{trial} - F^{old}$, and $C$ is a constant dependent on the specific target function for tuning the acceptance probability. The ultimate acceptance probability $p = \prod p^f$, where $p^f$ is the probability associated with each target function. If accepted, the configuration will be set as the starting configuration for the next swap trial; and if rejected, the configuration does not change.

(iii) Repeat the i) and ii) steps for a fixed number of rounds, or until being close to the target property.

## Data availability

All data needed to evaluate the conclusions in the paper are present in the paper and/or the Supplementary Materials. The raw data used in this work are available from the corresponding author upon reasonable request.

## Code availability

The Pytorch implementation of our GNN is available at https://github.com/Qi-max/gnn-for-glass.

## REFERENCES

1. Franceschetti, A. & Zunger, A. The inverse band-structure problem of finding an atomic configuration with given electronic properties. *Nature* **402**, 60–63 (1999).

2. Wiecha, P. R. *et al.* Evolutionary multi-objective optimization of colour pixels based on dielectric nanoantennas. *Nat. Nanotechnol.* **12**, 163–169 (2017).

3. d'Avezac, M., Luo, J.-W., Chanier, T. & Zunger, A. Genetic-Algorithm Discovery of a Direct-Gap and Optically Allowed Superstructure from Indirect-Gap Si and Ge Semiconductors. *Phys. Rev. Lett.* **108**, 27401 (2012).

4. Noé, F., Olsson, S., Köhler, J. & Wu, H. Boltzmann generators: Sampling equilibrium states of many-body systems with deep learning. *Science (80-. ).* **365**, 982–983 (2019).

5. Wiecha, P., Arbouet, A., Girard, C. & Muskens, O. Deep learning in nano-photonics: inverse design and beyond. *Photonics Res.* (2021). doi:10.1364/prj.415960

6. Zunger, A. Inverse design in search of materials with target functionalities. *Nat. Rev. Chem.* **2**, 121 (2018).

7. Kim, B., Lee, S. & Kim, J. Inverse design of porous materials using artificial neural networks. *Sci. Adv.* **6**, eaax9324 (2020).

8. Yao, Z. *et al.* Inverse design of nanoporous crystalline reticular materials with deep generative models. *Nat. Mach. Intell.* **3**, 76–86 (2021).

9. Sun, Y., Concustell, A. & Greer, A. L. Thermomechanical processing of metallic glasses: extending the range of the glassy state. *Nat. Rev. Mater.* **1**, 16039 (2016).

10. Wang, W. H., Dong, C. & Shek, C. H. Bulk metallic glasses. *Materials Science and Engineering R: Reports* **44**, 45–90 (2004).

11. Ketov, S. V *et al.* Rejuvenation of metallic glasses by non-affine thermal strain. *Nature* **524**, 200–203 (2015).

12. Ninarello, A., Berthier, L. & Coslovich, D. Models and Algorithms for the Next Generation of Glass Transition Studies. *Phys. Rev. X* **7**, 21039 (2017).

13. Cubuk, E. D. *et al.* Identifying structural flow defects in disordered solids using machine-learning methods. *Phys. Rev. Lett.* **114**, 108001 (2015).

14. Wang, Q. & Jain, A. A transferable machine-learning framework linking interstice distribution and plastic heterogeneity in metallic glasses. *Nat. Commun.* **10**, 5537 (2019).

15. Schoenholz, S. S., Cubuk, E. D., Sussman, D. M., Kaxiras, E. & Liu, A. J. A structural approach to relaxation in glassy liquids. *Nat. Phys.* **12**, 469–471 (2016).

16. Wang, Q. *et al.* Predicting the propensity for thermally activated β events in metallic glasses via interpretable machine learning. *npj Comput. Mater.* **6**, 194 (2020).

17. Devlin, J., Chang, M.-W., Lee, K. & Toutanova, K. BERT: Pre-training of Deep Bidirectional Transformers for Language Understanding. in *Proceedings of the 2019 Conference of the North {A}merican Chapter of the Association for Computational Linguistics: Human Language Technologies, Volume 1 (Long and Short Papers)* 4171–4186 (Association for Computational Linguistics, 2019). doi:10.18653/v1/N19-1423

18. LeCun, Y., Bengio, Y. & Hinton, G. Deep learning. *Nature* **521**, 436–444 (2015).

19. Fan, Z. & Ma, E. Predicting orientation-dependent plastic susceptibility from static structure in amorphous solids via deep learning. *Nat. Commun.* **12**, 1506 (2021).

20. Bapst, V. *et al.* Unveiling the predictive power of static structure in glassy systems. *Nat. Phys.* **16**, 448–454 (2020).

21. Hamilton, W., Ying, Z. & Leskovec, J. Inductive Representation Learning on Large Graphs. in *Advances in Neural Information Processing Systems* (eds. Guyon, I. et al.) **30**, (Curran Associates, Inc., 2017).

22. Fout, A., Byrd, J., Shariat, B. & Ben-Hur, A. Protein Interface Prediction using Graph Convolutional Networks. in *Advances in Neural Information Processing Systems* (eds. Guyon, I. et al.) **30**, (Curran Associates, Inc., 2017).

23. Duvenaud, D. *et al.* Convolutional networks on graphs for learning molecular fingerprints. in *Advances in Neural Information Processing Systems* (2015).

24. Xie, T. & Grossman, J. C. Crystal Graph Convolutional Neural Networks for an Accurate and Interpretable Prediction of Material Properties. *Phys. Rev. Lett.* **120**, 145301 (2018).

25. Chen, C., Ye, W., Zuo, Y., Zheng, C. & Ong, S. P. Graph Networks as a Universal Machine Learning Framework for Molecules and Crystals. *Chem. Mater.* **31**, 3564–3572 (2019).

26. Wu, Z. *et al.* A Comprehensive Survey on Graph Neural Networks. *IEEE Trans. Neural Networks Learn. Syst.* **32**, 4–24 (2021).

27. He, K., Zhang, X., Ren, S. & Sun, J. Deep Residual Learning for Image Recognition. in *2016 IEEE Conference on Computer Vision and Pattern Recognition (CVPR)* 770–778 (2016). doi:10.1109/CVPR.2016.90

28. Vaswani, A. *et al.* Attention is all you need. in *Advances in Neural Information Processing Systems* (2017).

29. Veličković, P. *et al.* Graph attention networks. in *6th International Conference on Learning Representations, ICLR 2018 - Conference Track Proceedings* (2018).

30. Srivastava, N., Hinton, G., Krizhevsky, A., Sutskever, I. & Salakhutdinov, R. Dropout: A Simple Way to Prevent Neural Networks from Overfitting. *J. Mach. Learn. Res.* **15**, 1929–1958 (2014).

31. Vinyals, O., Bengio, S. & Kudlur, M. Order Matters: Sequence to sequence for sets. (2016).

32. Hochreiter, S. & Schmidhuber, J. Long Short-Term Memory. *Neural Comput.* **9**, 1735–1780

(1997).

33. Liu, Z. *et al.* GeniePath: Graph Neural Networks with Adaptive Receptive Paths. (2018).

34. Huang, G., Liu, Z., Maaten, L. Van Der & Weinberger, K. Q. Densely Connected Convolutional Networks. in *2017 IEEE Conference on Computer Vision and Pattern Recognition (CVPR)* 2261–2269 (2017). doi:10.1109/CVPR.2017.243

35. Paszke, A. *et al.* Automatic differentiation in PyTorch. in *NIPS-W* (2017).

36. Cheng, Y. Q. & Ma, E. Atomic-level structure and structure-property relationship in metallic glasses. *Prog. Mater. Sci.* **56**, 379–473 (2011).

37. Maloney, C. E. & Lemaître, A. Amorphous systems in athermal, quasistatic shear. *Phys. Rev. E* **74**, 16118 (2006).

38. Thompson, A. P., Swiler, L. P., Trott, C. R., Foiles, S. M. & Tucker, G. J. Spectral neighbor analysis method for automated generation of quantum-accurate interatomic potentials. *J. Comput. Phys.* **285**, 316–330 (2015).

39. Cheng, Y. Q., Ma, E. & Sheng, H. W. Atomic Level Structure in Multicomponent Bulk Metallic Glass. *Phys. Rev. Lett.* **102**, 245501 (2009).

40. Hastings, W. K. Monte Carlo Sampling Methods using Markov Chains and their Applications. *Biometrika* **57**, 97–109 (1970).

**Acknowledgements:**

This work is supported by the Foundation from Institute of Materials, China Academy of Engineering Physics (Grant No. TP02201713). Part of the simulations are performed on the Maryland Advanced Research Computing Center (MARCC).

**Author contributions:**

Q. W. and L. F. Z. initiated the project. Q. W. designed the model architecture. Q. W. and L. F. Z. trained the model and conducted the analyses. Q. W. conducted the deformation simulations. Q. W. wrote the paper, and both authors commented on the manuscript.

**Competing interests:**

The authors declare that they have no competing interests.

# Supplementary Materials for
# Inverse design of glass structure with deep graph neural networks

Qi Wang[1 *], Longfei Zhang[2]

[1] Science and Technology on Surface Physics and Chemistry Laboratory, P. O. Box 9-35, Jiangyou, Sichuan 621908, China

[2] School of Software, Beihang University, Beijing 100191, China

*Corresponding author. Email: qiwang.mse@gmail.com

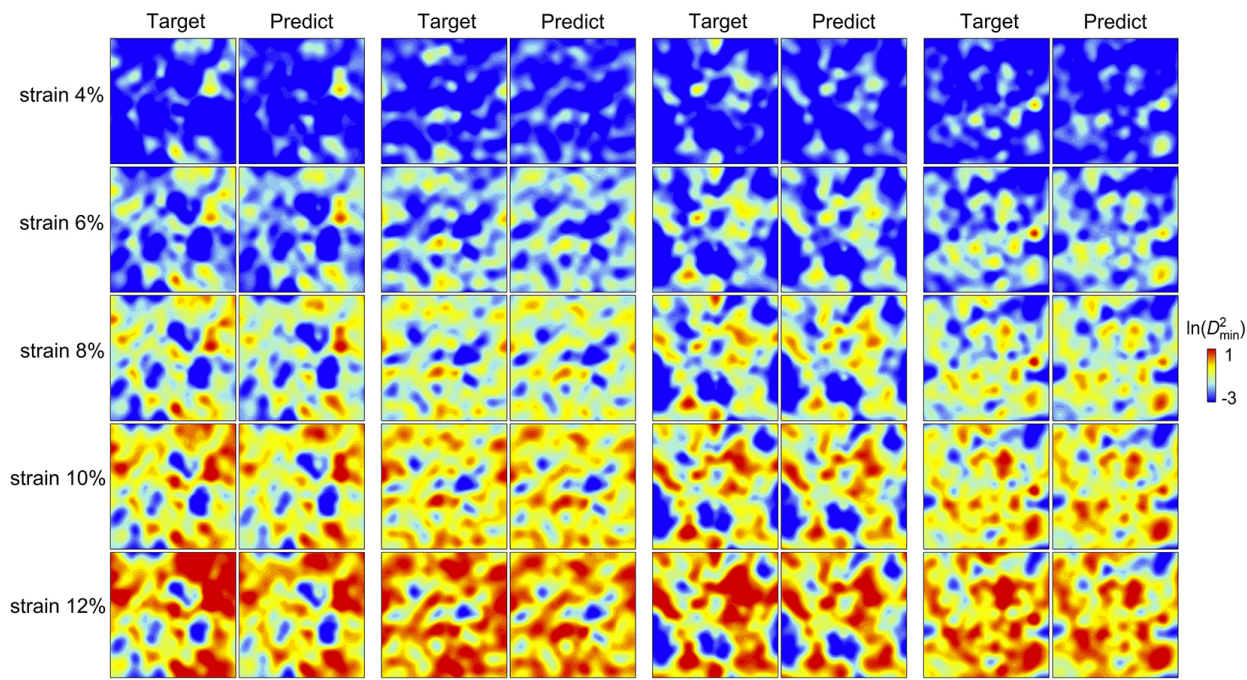

**Supplementary Figure 1.** Target versus GNN predicted $\ln(D_{min}^2)$ in typical slices of 3 Å thickness in a $Cu_{64}Zr_{36}$ – $10^9$ K s$^{-1}$ configuration at strains of 4%, 6%, 8%, 10% and 12%. The GNN model trained and verified in $Cu_{64}Zr_{36}$ – $10^{10}$ K s$^{-1}$ at each strain is used for prediction (generalization from $Cu_{64}Zr_{36}$ – $10^{10}$ K s$^{-1}$ to $Cu_{64}Zr_{36}$ – $10^9$ K s$^{-1}$).

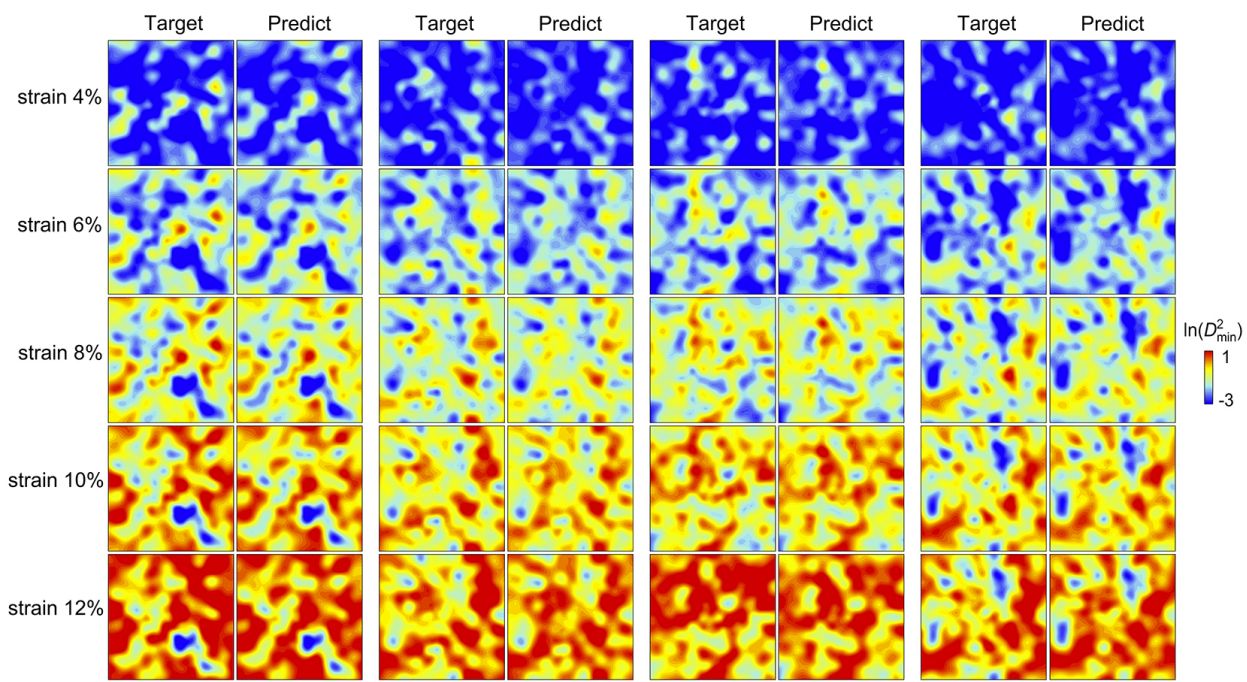

**Supplementary Figure 2.** Target versus GNN predicted $\ln(D^2_{\min})$ in typical slices of 3 Å thickness of a $Cu_{64}Zr_{36}$ – $10^{10}$ K s$^{-1}$ configuration at strains of 4%, 6%, 8%, 10% and 12%.

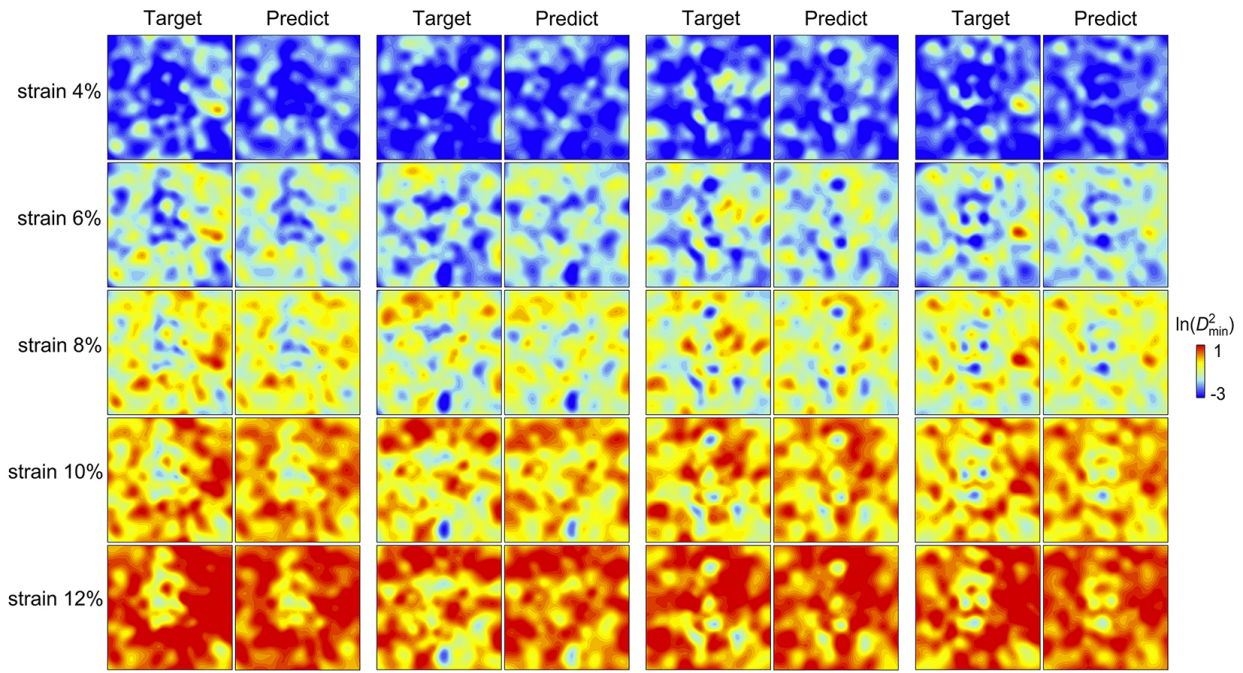

**Supplementary Figure 3.** Target versus GNN predicted $\ln(D^2_{min})$ in typical slices of 3 Å thickness of a $Cu_{50}Zr_{50}$ – $10^9$ K s$^{-1}$ configuration at strains of 4%, 6%, 8%, 10% and 12%. The GNN model trained and verified in $Cu_{50}Zr_{50}$ – $10^{10}$ K s$^{-1}$ at each strain is used for prediction (generalization from $Cu_{50}Zr_{50}$ – $10^{10}$ K s$^{-1}$ to $Cu_{50}Zr_{50}$ – $10^9$ K s$^{-1}$).

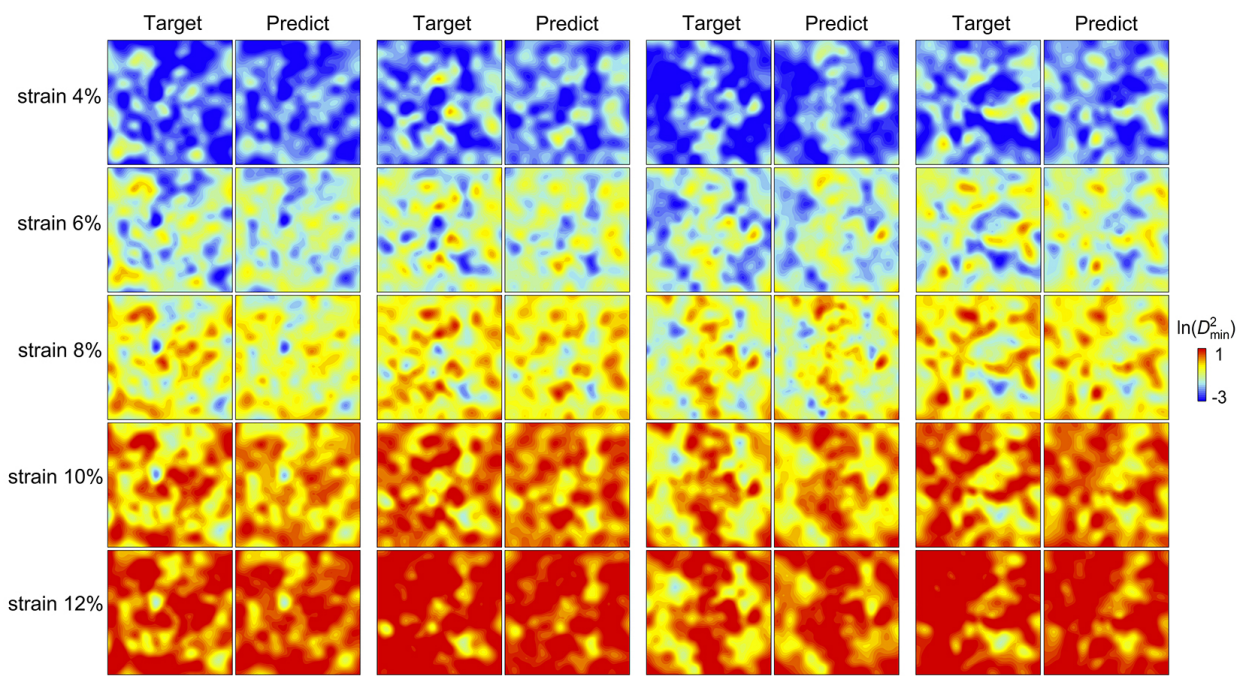

**Supplementary Figure 4.** Target versus GNN predicted $\ln(D^2_{min})$ in typical slices of 3 Å thickness of a $Cu_{50}Zr_{50}$ – $10^{10}$ K s$^{-1}$ configuration at strains of 4%, 6%, 8%, 10% and 12%.

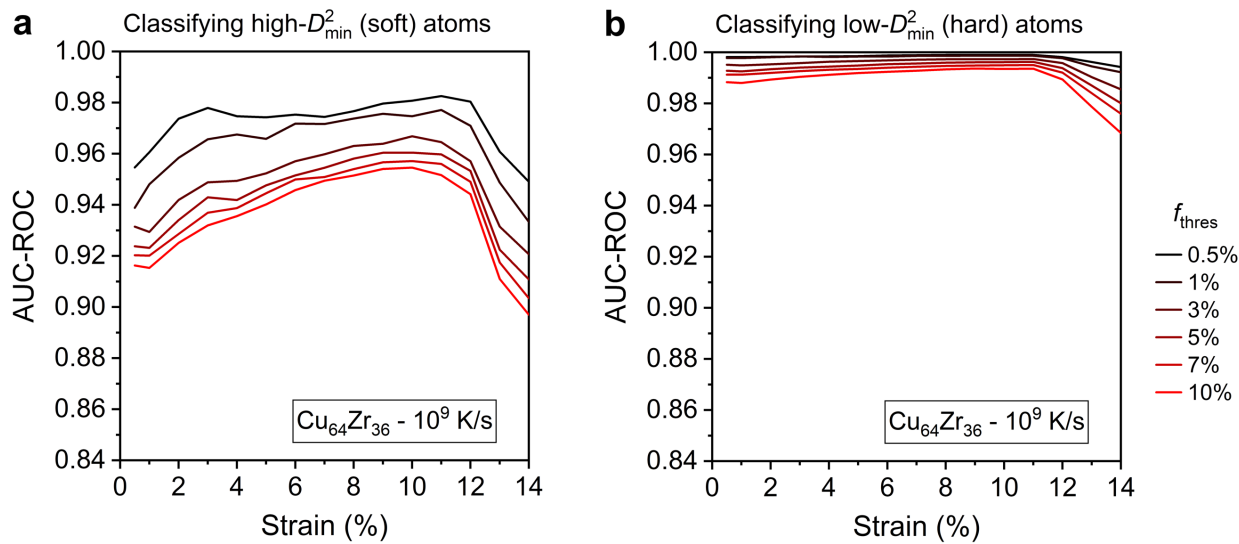

**Supplementary Figure 5.** Classifying the (**a**) soft and (**b**) hard atoms in $Cu_{64}Zr_{36} - 10^9$ K s$^{-1}$ test configurations using the GNN predicted $\ln(D^2_{min})$. The area under receiver operating characteristic curve (AUC-ROC) is derived as the classification metric (an AUC-ROC of 1.0 indicates perfect classification, 0.5 indicates random chance level). A series of fraction threshold, $f_{thres}$, are employed to setup the classification task ($f_{thres}$ = 0.5%, 1%, 3%, 7% and 10%), that is, atoms with $D^2_{min}$ among the highest (lowest) $f_{thres}$ at each strain will be designated as soft (hard) atoms, namely the positive class, and the remaining atoms will be the negative class.

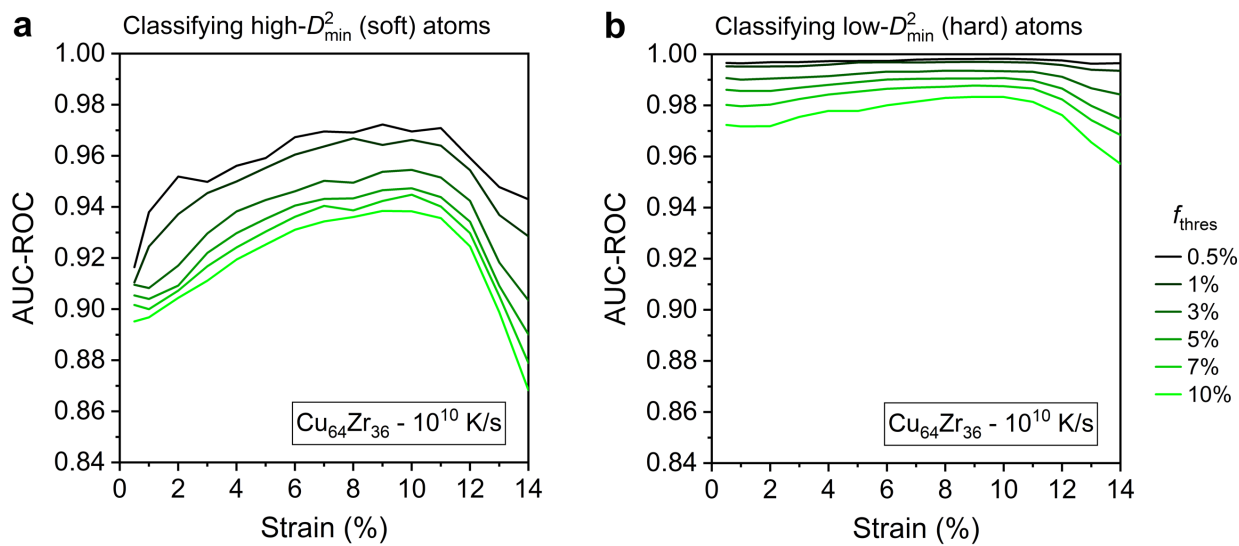

**Supplementary Figure 6.** Classifying the (**a**) soft and (**b**) hard atoms in $Cu_{64}Zr_{36} - 10^{10}$ K s$^{-1}$ test configurations using the GNN predicted $\ln(D^2_{min})$. The classification settings are the same as that described in Fig. S5.

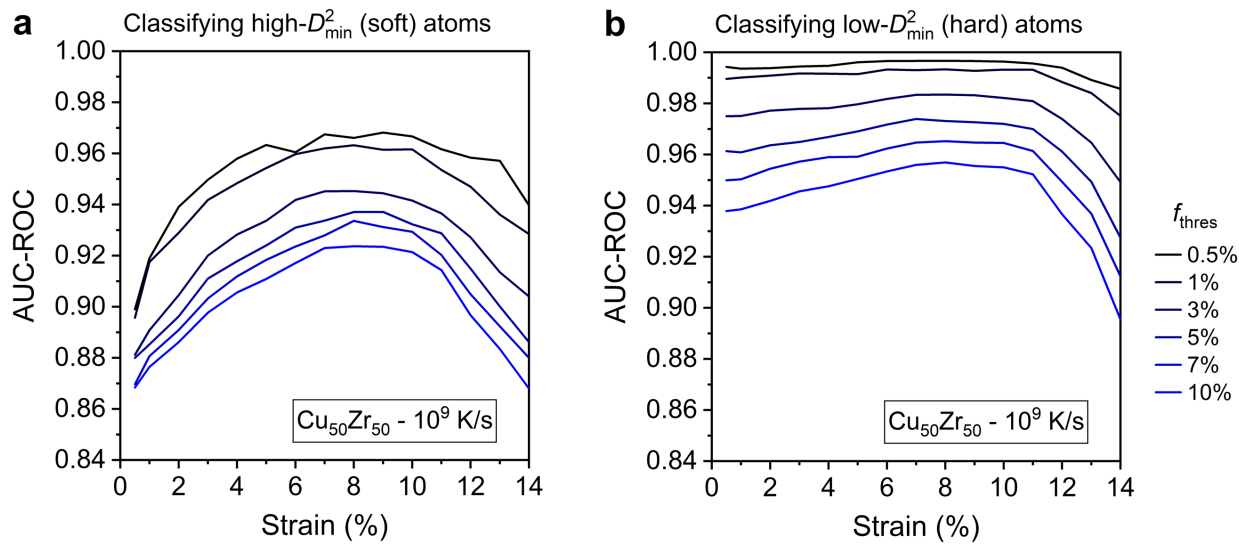

**Supplementary Figure 7.** Classifying the (**a**) soft and (**b**) hard atoms in $Cu_{50}Zr_{50}$ – $10^9$ K s$^{-1}$ test configurations using the GNN predicted $\ln(D^2_{min})$. The classification settings are the same as that described in Fig. S5.

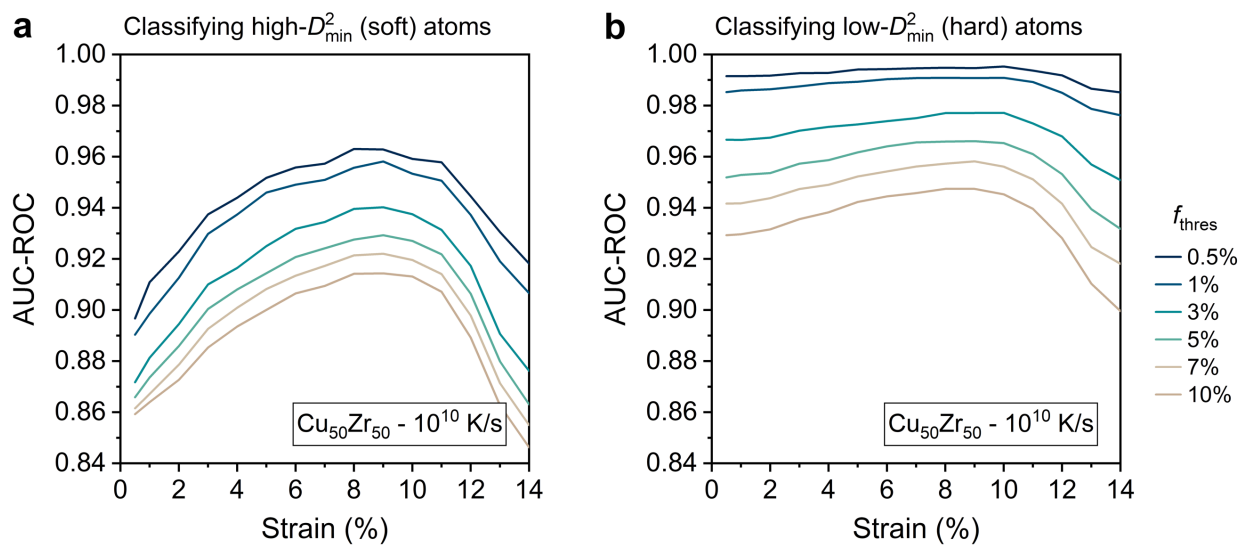

**Supplementary Figure 8.** Classifying the (**a**) soft and (**b**) hard atoms in $Cu_{50}Zr_{50}$ – $10^{10}$ K s$^{-1}$ test configurations using the GNN predicted $\ln(D^2_{min})$. The classification settings are the same as that described in Fig. S5.

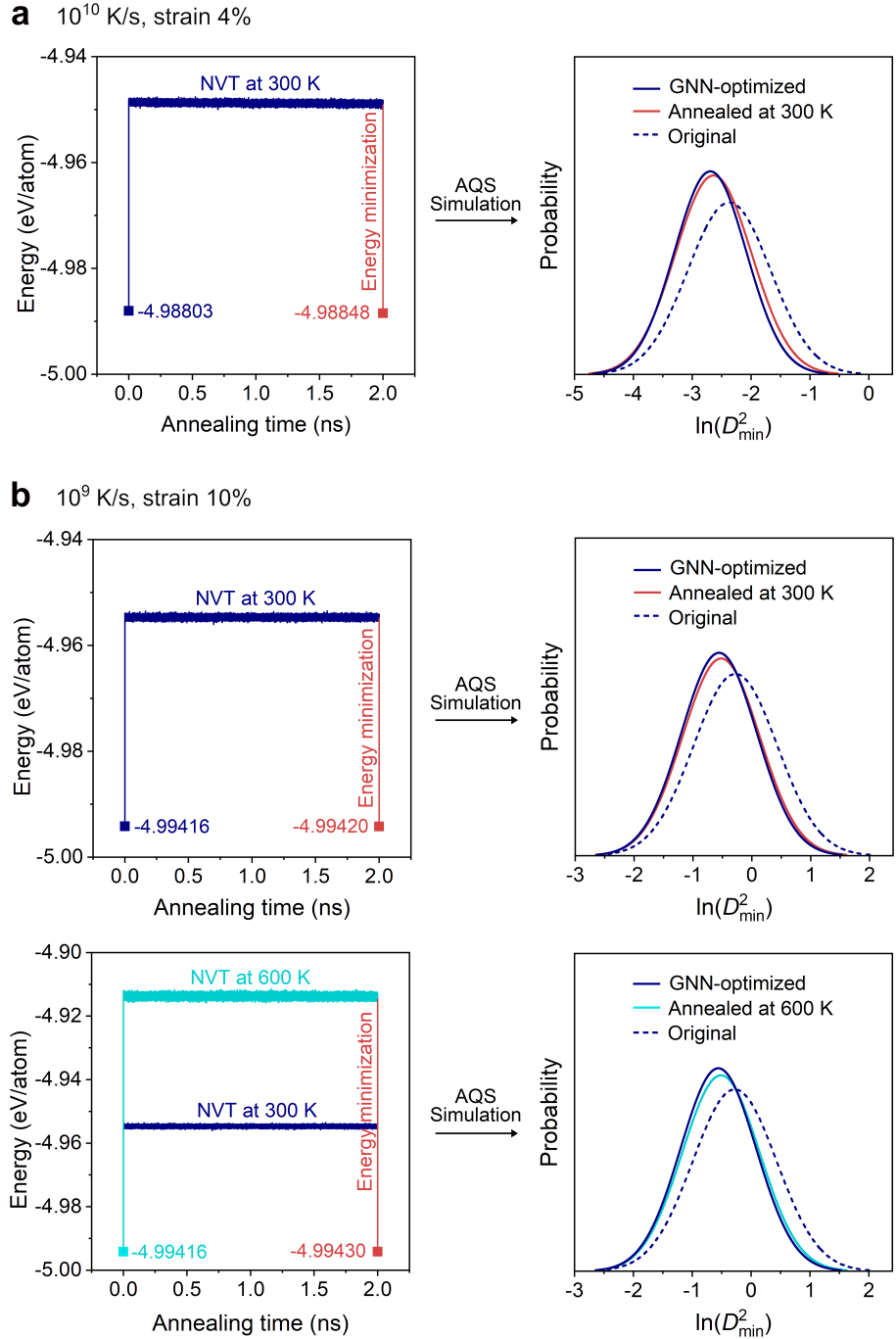

**Supplementary Figure 9.** (**a**) Annealing the GNN-optimized $Cu_{50}Zr_{50}$ – $10^{10}$ K s$^{-1}$ configuration under NVT ensemble at 300 K for 2 ns and get the inherent structure by energy minimization (left panel). Distribution of $\ln(D^2_{min})$ of the GNN-optimized configuration, the GNN-optimized configuration after annealing at 300 K, and the original unoptimized configuration, respectively (right). (**b**) Annealing the GNN-optimized $Cu_{50}Zr_{50}$ – $10^9$ K s$^{-1}$ configuration under NVT ensemble at 300 K or 600 K for 2 ns and get the inherent structure by energy minimization (left). Distribution of $\ln(D^2_{min})$ of the GNN-optimized configuration, the GNN-optimized configuration after annealing at 300 K or 600 K and the original unoptimized configuration, respectively (right).

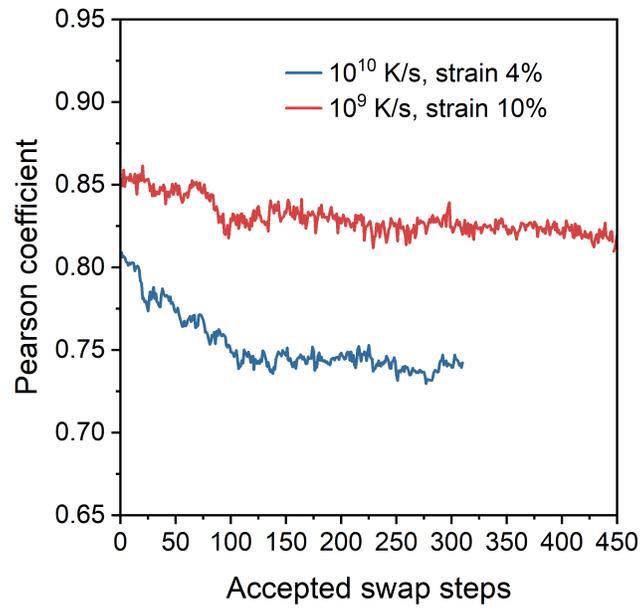

**Supplementary Figure 10.** The change of Pearson coefficient with the progress of optimization.